\documentclass[useAMS,usenatbib]{mn2e}

\usepackage{graphicx}

\title[GRB Cosmology]
{Gamma-Ray Bursts in $1.8 < z < 5.6$ Suggest that the Time
Variation of the Dark Energy is Small}
\author[Y. Kodama et al.]
{Yoshiki Kodama$^{1}$ 
\thanks{E-mail: kodama@astro.s.kanazawa-u.ac.jp (YK)}, 
Daisuke Yonetoku$^{1}$
\thanks{E-mail: yonetoku@astro.s.kanazawa-u.ac.jp (DY)}, 
Toshio Murakami$^{1}$,
Sachiko Tanabe$^{1}$, 
\newauthor   
Ryo Tsutsui$^{2}$ and Takashi Nakamura$^{2}$\\
$^{1}$Department of Physics, Faculty of Science, Kanazawa University,
Kakuma, Kanazawa, Ishikawa 920-1192, Japan\\
$^{2}$Department of Physics, Kyoto University,
Kyoto 606-8502, Japan\\
}
\begin{document}


\pagerange{\pageref{firstpage}--\pageref{lastpage}} \pubyear{2002}

\maketitle

\label{firstpage}

\begin{abstract}
We calibrated the peak energy-peak luminosity relation
of GRBs (so called Yonetoku relation) using 33~events with
the redshift $z < 1.62$ without assuming any  cosmological models. 
The luminosity distances to GRBs are estimated from those of 
large amount of Type Ia supernovae with $z<1.755$. This calibrated 
Yonetoku relation can be used as a new cosmic distance ladder 
toward higher redshifts. We determined the luminosity distances of 
30~GRBs in $1.8 < z < 5.6$ using the calibrated relation and 
plotted the likelihood contour in $(\Omega_m,\Omega_\Lambda)$ plane. 
We obtained $(\Omega_m, \Omega_{\Lambda})= (0.37^{+0.14}_{-0.11}, 
0.63^{+0.11}_{-0.14})$ for a flat universe. 
Since our method is free from the circularity problem, we can say 
that our universe in $1.8 < z < 5.6$ 
is compatible with the so called concordance cosmological model
derived for  $z < 1.8$.
This suggests that the time variation of the dark energy is
 small or zero up to $z\sim 6$.

\end{abstract}

\section{Introduction}
\label{sec:intro}

There are several distance indicators to determine the 
cosmological luminosity distance $d_L(z)$. If a certain distance 
indicator is calibrated without 
any cosmological models, the indicator can be used to determine 
the cosmological parameters such as $\Omega_m$, $\Omega_\Lambda$ 
and $w \equiv p/\rho$. In 1993, \citet{Phillips1993} discovered 
that the absolute magnitude of Type Ia supernovae at the peak is 
strongly correlated to the decline rate of the lightcurve after 
the maximum epoch. This correlation is rigidly calibrated using 
other distance indicators such as Tully-Fisher and
 Faber-Jackson relations which are free from 
any cosmological models. If the correlation does not depend 
on the redshift $z$, we can determine  $d_L(z)$ only from 
the observed maximum flux and the redshift
of the host galaxy.  Since the luminosity distance depends on 
the cosmological parameters, we can determine them from 
the estimated luminosity distances for high redshift Type Ia 
supernovae.

Using the calibrated correlation of Type Ia supernovae, 
the existence of the dark energy is strongly suggested first 
by \citet{Schmidt1998, Riess1998, Perlmutter1999}.
Thanks to the latest large number (41) of observations
 of Type Ia supernovae, the data with $z \le 1.755$ favor 
($\Omega_m, \Omega_\Lambda)=(0.27, 0.73)$ for a flat cosmology 
\citep{Riess2007}, which is usually called as the concordance 
cosmological model. However the most distant Type Ia supernova 
ever observed is at $z=1.755$ so that we need either further 
 Type Ia supernovae or other distance indicators 
to know the property of 
the dark energy beyond $z > 1.8$, while the anisotropy of 
the cosmic microwave background (CMB) gives us the information 
at $z=1089$ \citep{Spergel2007}.

One of the possible other distance indicators is 
Gamma-ray bursts (GRBs) whose maximum redshift observed is
 higher than that of Type Ia supernovae. 
At present the most distant 
GRB is at $z=6.3$ confirmed by the spectroscopic observation 
with Subaru telescope \citep{Kawai2006}. Since GRBs are known 
as the most violent and brightest explosion in the universe,
they might be a possible good distance indicator
 beyond $z > 1.8$. 
For this purpose, in part, several distance indicators have 
been proposed so far
\citep{Fenimore2000, Norris2000, Amati2002, Yonetoku2004,
Ghirlanda2004a, Liang2005, Firmani2006a}. 
Using these distance indicators, the cosmological parameters
are independently estimated by several authors
such as \citet{Ghirlanda2004b, Firmani2006b} and \citet{Schaefer2007}.

However the fundamental difficulty exists when we apply 
these distance indicators of GRBs to determine the cosmological 
parameters. At first, these indicators are established assuming 
the concordance cosmological model. After that, cosmological 
parameters are estimated by various methods. Therefore 
the same parameter set would be inevitably obtained. 
This logic is falling into a circularity problem.
Therefore, we should calibrate the distance indicators of GRBs 
without the theoretical concordance cosmology such as done for 
the Type Ia supernovae. Then we will suggest that the cosmic 
distance ladder is extended, and used for the measurement of 
the cosmological constant.

In 2004, \citet{Yonetoku2004} used 11 GRBs with known redshifts
($0.835 < z < 4.5$) at that time assuming a flat cosmology with
$\Omega_m=0.32,\Omega_\Lambda=0.68$ and 
$H_0=72~{\rm km~s^{-1}Mpc^{-1}}$ and derived that 
the peak luminosity ($L_p$) of GRBs correlates with the peak energy 
of the spectrum $E_p$ as $L_p\propto E_p^2$ (so called Yonetoku relation). 
The chance probability of this correlation is $5\times 10^{-9}$. 

In this Letter, using 63 known redshifts of  GRBs, we will 
try to determine the cosmological parameters 
at higher redshift ($ 1.8 < z < 5.6$) only with the observable
quantities such as the peak flux ($f_{p,obs}$) 
and the $E_{p,obs}$. 
Using the luminosity distances of many Type Ia supernovae
established without any cosmological models,
we will calibrate and reconstruct the Yonetoku relation for nearby 
33 GRBs for $z < 1.755$. 
 Applying the calibrated relation to 30 GRBs in $1.8 < z < 5.6$,
we determine their luminosity distances as a function of $z$.
Then we will plot the likelihood contour on the 
$(\Omega_m,\Omega_\Lambda)$ plane and obtain 
$\Omega_m=0.37^{+0.14}_{-0.11},
\Omega_{\Lambda}=0.63^{+0.11}_{-0.14}$ for a flat universe. 
This method (logic) is very similar to the calibration for 
Type Ia supernovae with the Tully-Fisher and the Faber-Jackson 
relations, and it is free from the circularity problem.
The essential point is that we use the luminosity distances 
of Type Ia supernovae like Tully-Fisher relation, 
which are well determined by the observations and are free from 
the cosmological models, to calibrate the Yonetoku relation.

In \S 2 we show how to calibrate the Yonetoku relation
using the  Type Ia supernovae.
In \S 3 we show the likelihood contour in 
$(\Omega_m,\Omega_\Lambda)$ plane and argue the cosmological 
parameters. \S 4 is devoted to discussions.
Throughout the paper, we adopt $H_0=70~{\rm km~s^{-1}Mpc^{-1}}$

\section{Calibrating Yonetoku relation}
Generally, the spectrum of the prompt emission of GRBs can be 
explained as an exponentially connected broken power-law model, 
so called Band function \citep{Band1993}. We adopted this model 
to the time averaged gamma-ray spectra of known redshift 
samples \citep{Yonetoku2004}. 
Then we can determine a peak energy, $E_{p}$, which corresponds 
to the energy at the maximum flux in $\nu F_{\nu}$ spectra. 
\citet{Yonetoku2004} found a strong correlation between 
the $E_{p}$ and the 1-second peak luminosity ($L_{p}$). 
Here, we estimate the $L_{p}$ in 1--$10^{4}$~{\rm keV} energy 
range in the rest frame of GRB. We used the data observed by 
several independent missions and instruments, so we included 
appropriate k-correction suggested  by \cite{Bloom2001} when we 
estimate the luminosity for each event. We used 33 GRBs 
for $z < 1.755$ as the calibration of the $E_{p}$--$L_{p}$ 
relation. 

\begin{figure}
\rotatebox{0}{\includegraphics[width=90mm]{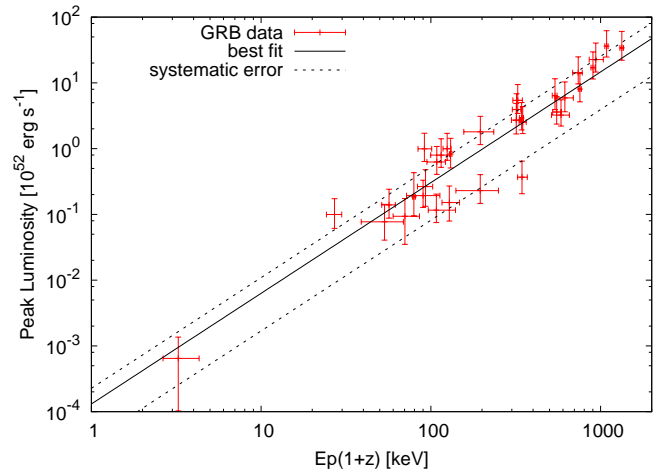}}
 \caption{The peak luminosity ($L_p$) and the peak energy
($E_p$) in the spectrum of 33~GRBs with $z < 1.62$. 
The linear correlation coefficient is 0.9478 and 
the chance probability is  $6.0 \times 10^{-17}$. 
The solid line is the best fit curve of 
${L_p}/{10^{52}{\rm erg~s^{-1}}} = 
1.31 \times 10^{-4}[E_p(1+z)/{\rm 1keV}]^{1.68}$ 
while two dashed lines are the curves including 
the systematic error. See text for details.}
\label{fig1}
\end{figure}

We found the empirical formula of the luminosity distance
of Type Ia supernovae in \citet{Riess2007} with 
$0.359 < z < 1.755$ as
\begin{eqnarray}
\frac{d_L(z)}{10^{27}{\rm cm}} =
14.57 \times z^{1.02} + 7.16 \times z^{1.76}.
\end{eqnarray}
This formula agrees with the real data within a relative
error of 1~\%. Here we note that this formula is not unique 
and the other formula is possible.
What is important here is  that we do not assume any 
cosmological models at this stage but simply assume that
the Type Ia supernovae are the standard candle for
$0.359 < z <1.755$ irrespective of the cosmological model.
 We apply this formula to 33 GRBs with
the redshift $z < 1.62$ in our sample to obtain
$L_p = 4\pi d_L(z)^2 f_{p,obs}$ while $E_p = (1+z) E_{p,obs}$.
In figure~\ref{fig1}, we show the calibrated $E_{p}$--$L_{p}$ 
relation for 33 GRBs within $z < 1.62$. The linear correlation 
coefficient is 0.9478 and the chance probability 
is $6.12 \times 10^{-17}$. 
We tried to find the best fit curve in the form as
\begin{eqnarray}
\label{eq:ep-l}
(\frac{L_p}{10^{52}{\rm erg~s^{-1}}})
= (1.31 \pm 0.67)\times 10^{-4} 
(\frac{E_p}{\rm 1keV})^{1.68 \pm 0.09}
\end{eqnarray}
In this equation, the error is expressed as the statistical 
uncertainty. However the data distribution has a larger 
deviation around the best fit line compared with the
expected Gaussian distribution. We estimated this systematic 
deviation in the normalization as $9.57 \times 10^{-5}$. 
The solid line is the best fit curve and two dashed lines 
are the curves including the systematic error in the 
normalization. 

\begin{figure}
\rotatebox{270}{\includegraphics[width=60mm]{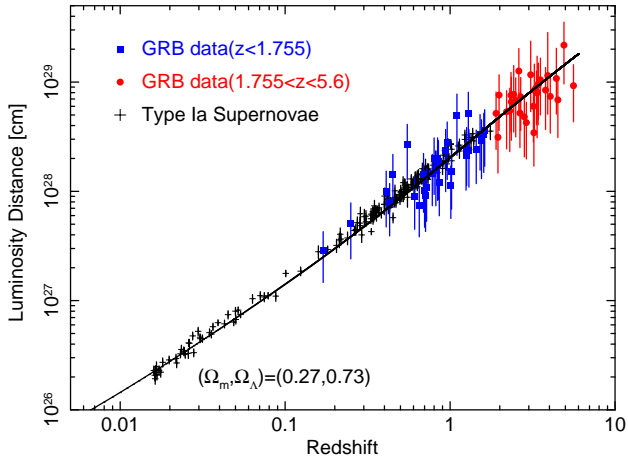}}
 \caption{The luminosity distance as a function of the
redshift measured by the calibrated $E_{p}$--$L_{p}$ relation.
The blue and the red points are the luminosity distance
of $z \le 1.755$ and $z > 1.755$, respectively.
The data of Type Ia supernovae are also plotted 
as the black cross points.
The uncertainty bar of each point includes the systematic
dispersion around the best fit line of figure~\ref{fig1}.
For the purpose of easy comparison, a solid line of the 
concordance model is also drawn.}
\label{fig2}
\end{figure}

\section{Cosmological Parameters}
At present, there are more than 100~GRBs with known redshift, 
but the $E_{p}$ has not been measured for all GRBs. 
Moreover the calibrated range in figure~\ref{fig1}
is $3.3 \le E_{p} \le 1333$~keV 
in the rest frame of GRBs, so we should use GRBs within 
this range to estimate the luminosity distance of $z > 1.755$.
Although we know the redshift of GRB~050904 is $z = 6.3$, 
its peak energy is reported as $E_{p} \sim 3~{\rm MeV}$. 
This is out of our calibration, so we exclude it from 
our sample. 
Then, we apply the calibrated Yonetoku relation to 30~GRBs
in $1.8 < z < 5.6$ to determine the luminosity distance as 
a function of $z$. In figure 2 we show the luminosity 
distance of 33+30 GRBs as a function of the redshift.
The blue and the red points are the luminosity distance
of $z \le 1.755$ and $z > 1.755$, respectively.
The uncertainty bar of each red point includes the systematic
dispersion around the best fit line of figure~\ref{fig1}

\begin{figure}
\includegraphics[width=84mm]{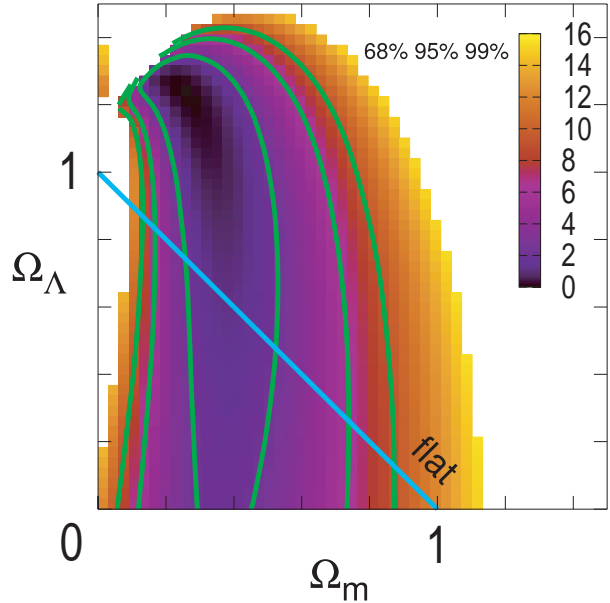}
\caption{The contour of the likelihood $\Delta \chi^2$ 
(see equation~\ref{eq:contour}) for the luminosity 
distances of 30~GRBs in $1.8 < z < 5.6$.
The significance levels of 68\%, 95\% and 99\%
are also shown on the same panel.
Compared with  the case of Type Ia supernovae, 
the shape of the contour stands vertical since 
the mean redshift is much larger. 
The most likelihood value of cosmological parameters are
$(\Omega_m, \Omega_\Lambda) = 
(0.25^{+0.27}_{-0.14}, 1.25^{+0.10}_{-1.25})$
while for a flat cosmology prior they are
$(\Omega_m, \Omega_\Lambda) = 
(0.37^{+0.14}_{-0.11}, 0.63^{+0.11}_{-0.14})$.}
\label{fig3}
\end{figure}

For each GRB with $z=z^i$ we have 
the observed peak flux($f_{p,obs}^i$) in the unit of 
${\rm erg~cm^{-2}s^{-1}}$ and the observed $E_{p,obs}^{i}$ 
in the unit of 1~keV. Then using the equation~\ref{eq:ep-l},
the luminosity distance can be derived as
\begin{eqnarray}
 d_L({z^i})=10^{24}{\rm cm}\sqrt{\frac{1.31}{4\pi f_{p,obs}^i}}
[{E_{p,obs}^{i}}(1+z^i)]^{1.68/2}.
\end{eqnarray}
In the $\Lambda$CDM-cosmology with
$\Omega_k \equiv \Omega_m + \Omega_\Lambda - 1$,
the luminosity distance is given by 
\begin{eqnarray}
{d_L^{th}}(z,\Omega_m,\Omega_\Lambda)&=&\left\{
\begin{array}{ll}
\frac{c}{H_0\sqrt{\Omega_k}}\sin(\sqrt{\Omega_k}F(z))
& \mbox{if}~~\Omega_k > 0\\
\frac{c}{H_0\sqrt{-\Omega_k}}\sinh(\sqrt{-\Omega_k}F(z))
& \mbox{if}~~\Omega_k < 0\\
\frac{c}{H_0}F(z)
& \mbox{if}~~ \Omega_k = 0\\
\end{array}
\right.\\
\mbox{where}~~F(z)&=&\int_0^z\frac{dz}{\sqrt{\Omega_m(1+z)^3
+\Omega_\Lambda-\Omega_k(1+z)^2}}.
\end{eqnarray}
We define a likelihood function as 
\begin{equation}
\label{eq:contour}
\Delta \chi^2 = \sum_{i=1}^{30}(\frac{\log d_L({z^i})-
\log {d_L^{th}}(z^i,\Omega_m,\Omega_\Lambda)}{\Delta d_L(z^i)})^2
- \chi_{best}^2.
\end{equation}
Here $\chi_{best}^{2}$ means the chi-square value for the
best fit parameter set of $\Omega_m$ and $\Omega_\Lambda$.
In figure~\ref{fig3} we show the contour of the likelihood 
$\Delta \chi^2$ for the luminosity distances of 30~GRBs 
in $1.8 < z < 5.6$. Compared with the case of Type Ia 
supernovae, the shape of the probability contour stands 
vertical since the luminosity distance strongly depends 
on $\Omega_m$ rather than $\Omega_\Lambda$ for higher 
redshift samples. This is clear from the functional form 
of $F(z)$. Without any prior the most likelihood value is 
$(\Omega_m, \Omega_\Lambda) = 
(0.25^{+0.27}_{-0.14}, 1.25^{+0.10}_{-1.25})$
while for a flat cosmology prior that is 
$(\Omega_m, \Omega_\Lambda) 
= (0.37^{+0.14}_{-0.11}, 0.63^{+0.11}_{-0.14})$
with $1~\sigma$ uncertainty.

\begin{figure}
\includegraphics[width=84mm]{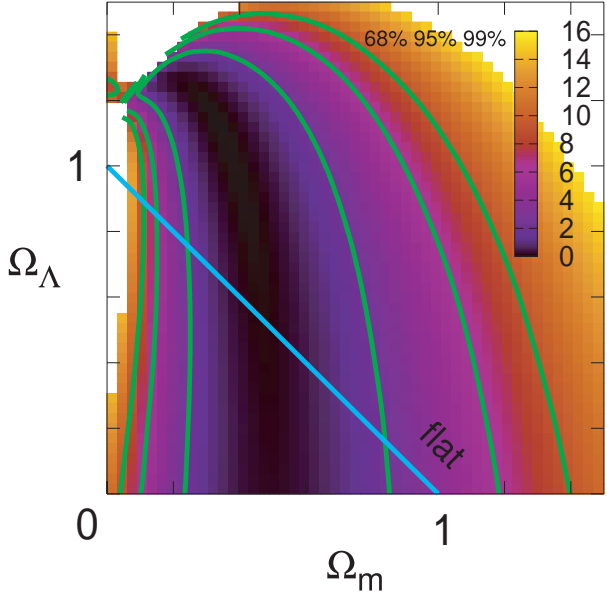}
 \caption{The same as fig.~\ref{fig3} but for the luminosity 
distances of 16~GRBs in $3 < z < 5.6$. The shape of the contour 
stands more vertical than fig.~\ref{fig2}. 
The most likelihood values are 
$(\Omega_m, \Omega_\Lambda) = 
(0.33^{+0.52}_{-0.26}, 1.14^{+0.21}_{-1.14})$
while for a flat cosmology prior they are
$(\Omega_m, \Omega_\Lambda) = 
(0.49^{+0.33}_{-0.24}, 0.53^{+0.22}_{-0.37})$.}
\label{fig4}
\end{figure}

In figure~\ref{fig4}, we show the same contour as figure~\ref{fig3} 
but for the luminosity distances of 16~GRBs in $3 < z < 5.6$. 
The shape of the contour stands more vertical than 
fig.~\ref{fig3}, which should be so. 
Without any prior the most likelihood values of cosmological 
parameters are $(\Omega_m, \Omega_\Lambda) = 
(0.33^{+0.52}_{-0.26}, 1.14^{+0.21}_{-1.14})$ 
while for a flat cosmology prior they are $(\Omega_m, \Omega_\Lambda) =
(0.49^{+0.33}_{-0.24}, 0.53^{+0.22}_{-0.37})$. We see for a flat
cosmology the value of  $(\Omega_m, \Omega_\Lambda)$ for higher
redshift samples is compatible with that for whole samples if
we take the error  into account.

\section{Discussions}

In this Letter we extended the cosmic distance ladder 
using GRBs calibrated by Type Ia supernovae, and argued 
the cosmological parameters using only the GRBs in 
$1.8 < z < 5.6$ except for a flat cosmology prior. 
Since our region of $1.8 < z < 5.6$ has not been 
explored yet, this is the first report to estimate 
the cosmological parameters up to $z = 5.6$.
It is possible to combine our data with 
(1) Type Ia supernova, (2) CMB, (3) Baryon Acoustic Oscillation 
(4) the large scale structure measurement and
(5)weak gravitational lensing, to constrain 
the $w(z) \equiv p/\rho$ parameter of the equation of state 
for the dark energy (Tsutsui et al.). 

The calibrated Yonetoku relation has a large dispersion
in the normalization which is mainly caused by the 
systematic error. Therefore, currently, the measurement of
the luminosity distance is not so accurate compared with
the other distance indicators such as Type Ia supernova.
 It may be difficult to 
discuss the detailed time history of the cosmological 
parameters yet. However, if this deviation is the intrinsic 
property of GRBs, we will be able to discover the hidden 
physical quantities, or distinguish a possible sub-class 
from entire population of GRBs like Type Ia supernova.
 These improvements in
the cosmic distance ladder will lead us to explore 
the deep space with better accuracy in near future. 

As for the Amati-relation \citep{Amati2002} which relates 
the total isotropic energy to $E_p$, possible selection 
bias effects and evolution effects are claimed 
\citep{Li2007, Butler2007} while \citet{Willingale2007} 
argues against such effects. It is important to check 
possible selection bias and evolution effects in 
the Yonetoku relation also.
\citet{Tanabe2008} examined the evolution effect as well as 
the observational selection bias assuming the concordance 
cosmology, and found that they are quite small. We think that 
the selection bias and evolution effects are not larger than 
the systematic uncertainty in the normalization of 
the calibrated Yonetoku relation, although we agree that 
further examinations are required. 

Theoretical models of dark energy are reviewed, for example, 
in a recent paper by \citet{Frieman2008}.
In some models such as scalar field models, dark energy 
looks like the cosmological constant for low redshift 
$z <2$ but not for high $z >2 $. Therefore the estimate of 
the cosmological parameters for $z >2 $ is important 
either to refute or confirm such models. 
Our present result for $1.8 < z < 5.6$ gives 
$(\Omega_m, \Omega_\Lambda) = 
(0.37^{+0.14}_{-0.11}, 0.63^{+0.11}_{-0.14})$
for a flat cosmology prior while Type Ia supernova
for $z < 1.755$ does $(\Omega_m, \Omega_\Lambda) =(0.27,0.73)$ 
(\cite{Riess2007}), which means that  $(\Omega_m, \Omega_\Lambda)$ 
for $z < 1.8$ and for $1.8 < z < 5.6$ are the same within 
the $1~sigma$ statistical error. This suggests that 
$(\Omega_m, \Omega_\Lambda)$ did not change so much from $z=5.6$ 
to $z=0$ so that such a scalar field¡¡model should look like
the cosmological constant up to $z \sim 6$ at least.

\section*{Acknowledgments}
This work is supported in part by the Grant-in-Aid from the 
Ministry of Education, Culture, Sports, Science and Technology
(MEXT) of Japan,  No.19540283,No.19047004, No.19035006(TN),
and  No.18684007 (DY).

%



\label{lastpage}

\end{document}